\documentclass[12pt]{iopart}
\usepackage{epsf,rotate}
\usepackage{graphicx}
\begin{document}

\title[Characterizing Synchronization using Symbolic Representations]{Characterizing Synchronization in Time Series using Information Measures Extracted from Symbolic Representations}

\author{Roberto Monetti, Wolfram Bunk, Ferdinand Jamitzky}

\address{Max-Planck-Institut f\"ur extraterrestrische Physik, Giessenbachstr. 1, 85748 Garching, Germany}
\ead{monetti@mpe.mpg.de}
\begin{abstract}
We present a methodology to characterize synchronization in time
series based on symbolic representations. A symbol is linked to a
sequence of numbers through the rank-order of its values. A
representation of a time series results after mapping all sequences into
symbols. We propose a transcription scheme between symbolic representations to
study the dynamics of coupled systems. This scheme allows us to use
elements of group theory and to derive information measures to assess the degree of synchronization.
We apply our method to a prototype non-linear system which displays a
rich coupled dynamics. 
\end{abstract}
\maketitle

\section{Introduction}
Synchronization phenomena are ubiquitous in Nature. They take place among coupled oscillatory systems. Its occurrence is not restricted to periodic systems but it is also observed in non-linear chaotic systems. In this case, its emergence is by no means trivial due to the high sensitivity of chaotic systems to initial conditions.
Examples of synchronization arise in different fields of science like electronics (e.g. coupled circuits), physiology (e.g. between cardiac and respiratory systems or EEG signals) \cite{cr,eeg}, extended ecological systems \cite{eco} or in non-linear optics (e.g. coupled laser systems with feedback). Different synchronization states have been identified in the study of coupled chaotic systems, namely complete synchronization \cite{cs}, phase \cite{ps1,ps2} and lag synchronization \cite{ls}, generalized synchronization \cite{gs1,gs2}, etc (for a review about synchronization in chaotic systems see \cite{boca}). 
\begin{figure}
\centering
\includegraphics[width=13.cm,angle=0]{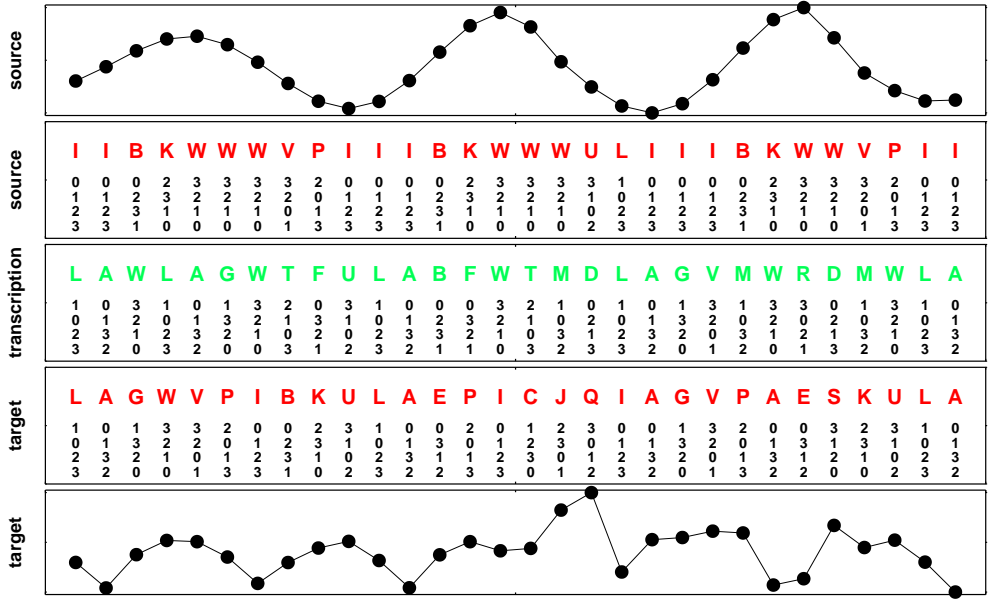}
%
%
\caption{Example of symbolic representations of time series for sequence length $p=4$. Symbols in red correspond to the representations and the green symbols indicate the transcriptions that have to be applied to the upper symbols (source) to obtain the lower ones (target). Note that this operation is not commutative.}
\label{fig:1}       
\end{figure}
Here, we present a methodology to characterize synchronization in coupled systems where information measures are obtained using symbolic representations of time series. 
\section{Method}
Let $x$ be a time series and $q=(x_0, \ldots ,x_{p-1})$ be a sequence of length $p$ extracted from $x$. The symbol $Q$ associated to $q$ is defined as the rank-ordered indices of the components of $q$.
For instance, for $q=(1.6,1.3,1.4,1.5)$, the symbol associated to $q$ is $Q=(3,0,1,2)$. This symbolic representation was first introduced by Bandt et al. \cite{pe} in the context of complexity analysis of time series. This approach motivated some studies of the characterization of similarities in time series. For the interested reader see \cite{liu,witt}. It should be mentioned that the occurrence of identical values in $q$ has not been considered. When the sequence contains equal values, one can always add a small random perturbation to avoid this case.

\begin{figure}
\centering
\includegraphics[width=13.cm,angle=0]{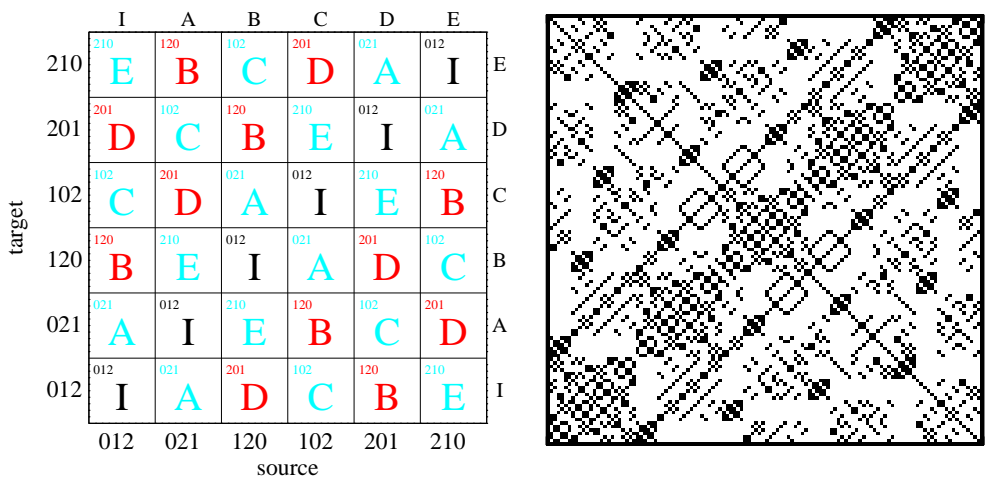}
%
%
\caption{Left: Transcription matrix for sequence length $p=3$. The blue (red) symbols belong to order 2 (3) class, respectively. The identity symbol (black) is a singular one symbol class satisfying $T^n = I, \; \forall n$. Source symbols are displayed at the bottom and target symbols on the left of the transcription matrix. Right: For $p=5$ the transcription matrix has $(5!)^2$ elements. This matrix shows the positions of the elements belonging to order 2 class, i.e. the structure generated by order 2 transcriptions for $p=5$.}
\label{fig:2}       
\end{figure}

Figure 1 shows symbolic representations of two time series (red symbols) for $p=4$. Given two symbols $A_1$ and $A_2$ there always exists a symbol $T$, in the following called transcription, such that $T[A_1]= A_2$. The action of symbol $T$ is defined as follows. Let $A_1=(j_0,j_1, \ldots , j_{p-1})$ and $T=(k_0,k_1, \ldots , k_{p-1})$. Then,
\begin{equation}
T[A_1]=(j_{k_0}, j_{k_1}, \ldots , j_{k_{p-1}}).
\end{equation}
It should be noted that the set of symbols form a finite non-Abelian group of order $p!$ with operation $T$ known as the symmetric group $S_p$.
Green symbols in Fig. 1 indicate the transcriptions between the symbolic representations of the time series.
The group $S_p$ can be partitioned into non-overlapping classes
$\mathcal{C}_i$ $(S_p= \bigcup \mathcal{C}_i)$ satisfying a power relation,
namely if $T \in \mathcal{C}_n$ then $T^n =I$, where $I=(0,1, \ldots , p-1)$
is the identity symbol and $T^n$ is the composition $T[T^{n-1}]$ with $n \ge 1$ and $T^0 \equiv I$. Figure 2 (left panel) shows the transcription matrix
for $p=3$, where the three existing order classes, i.e. $T = I$ (black symbol), $T^2 = I$ (blue symbols), and $T^3 =I$ (red symbols) are shown.
It is worth discussing the action of transcriptions for different order classes. The identity transcription leaves symbols unchanged thus it is the simplest transcription. For $p=3$, consider transcription $A=(0,2,1)$  which belongs to order 2 class and apply it to $E=(2,1,0)$ (see Fig. \ref{fig:2}).
\begin{equation}
A[E]=D=(2,0,1).
\end{equation}
Then, the action of $A$ is identical to one transposition, i.e. the interchange of $0$ and $1$ in symbol $E$. However, if we consider $B=(1,2,0)$ that belongs to order 3 class and apply it to $E=(2,1,0)$ the result is $C=(1,0,2)$. We have to perform either two transpositions or one cyclic permutation on $E$ to obtain $C$. 
Note that for $p=3$ all order 2 transcriptions cause a one transposition change while all order 3 transcriptions lead to two transpositions change. 
Thus, we interpret order 3 transcriptions as "more complex" than order 2 transcriptions. For longer sequences, we can still identify order classes in term of the action of their component symbols although the description becomes more difficult. From this point of view, the order of a class offers a rough estimation of the "complexity" of the transcription. 

The order classes satisfy an important property of invariance. Let $A$ and $B$ two symbols connected by the transcription $T$, i.e. $T[A]=B$ and suppose that $T^N = I$, i.e. $T \in \mathcal{C}_N$. Let $Y$ an arbitrary transcription such that $Y[A]=C$ and $Y[B]=D$. There always exists a transcription $T'$ such that $T'[C]=D$. We will prove that $T'$ belongs to order $N$ class as well. In fact, 
\begin{equation}
T'[C]=D \; \Rightarrow \; T'[Y[A]]=Y[B].
\label{eq3}
\end{equation}
If we apply $Y^{-1}$ on the left to both sides of Eq. \ref{eq3} we obtain
\begin{equation}
Y^{-1}T'[Y[A]]=B ,
\label{eq4}
\end{equation}
which implies
\begin{equation}
T=Y^{-1}T'Y.
\label{eq5}
\end{equation}
Since $T \in C_N$ we obtain
\begin{equation}
Y^{-1}T'^{N}Y=I,
\label{eq6}
\end{equation}
so $T$ and $T'$ belong to the same order class. This property of invariance also implies that $T$ and $T^{-1}$ belong to the same order class. However, an order class is not a group since it does not satisfy closure. Note that Eq. (\ref{eq5}) implies that order classes are also conjugacy classes. Figure 2 (right panel) shows the structure generated by the set $\mathcal{C}_2$ for $p=5$. The symmetry displayed by this structure is a general property found in all order classes since it is a consequence of Eq. \ref{eq6}.

The action of a transcription is just equivalent to applying permutations. It is well known that any permutation can be written as a product of disjoint cyclic permutations (DCP). Using this fact, one can prove that the order of any transcription is the least common multiple (LCM) of the lengths of the DCP. Since the sum of the lengths of the DCP equals the sequence length $p$, the succession of order classes is never interrupted up to order $p$. 
For $p \ge 7$, gaps of missing order classes always appear. For example for $p=7$, order 8, 9, and 11 classes are missing since there is no possible splitting of a sequence of length 7 in DCP which satisfy the LCM condition. However, order 10 and 12 classes are present since one can have a combination of DCP of lengths 2 and 5 for order 10 and 3 and 4 for order 12.

We now focus on the probability density of transcriptions.
Consider a source and a target symbolic representations generated by the actual coupled dynamics of the time series. Given a sequence of length $p$, the set of all feasible symbols $\mathcal{S}^{1}= \{ X_i \}$ and $\mathcal{S}^{2}= \{ X_j \}$ conform the state spaces for the source and the target representations, respectively. The probability density of transcriptions $P_{T}(p)$ can be written as follows
\begin{equation}
P_{T_k}(p) = \sum_{\Omega=\{(i,j): \; T_k[X_i] = X_j \}} P^{C}(X_i,X_j),
\label{eq7}
\end{equation}
where $X_i \in \mathcal{S}^{1}$, $X_j \in \mathcal{S}^{2}$, and $P^{C}(X_i,X_j)$ is the joint probability density. 
Let $P^{(1)} (X_i)$ and $P^{(2)} (X_j)$ be the marginal probability densities of the symbols $X_i$ and $X_j$ in state spaces $\mathcal{S}^{1}$ and $\mathcal{S}^{2}$, respectively. 
The matrix $M_{i,j}=P^{(1)} (X_i) P^{(2)} (X_j)$ is the probability density
matrix of transcriptions for two independent processes. 
In this case, the probability density of transcriptions $P^{e}_{T}(p)$ can be evaluated as follows
\begin{equation}
P^{e}_{T_k}(p) = \sum_{\Omega=\{(i,j): \; T_k[X_i] = X_j \}} M_{i,j}, 
\label{eq8}
\end{equation}
where $X_i \in \mathcal{S}^{1}$ and $X_j \in \mathcal{S}^{2}$.
The aim is to find an information measure to assess how much $P_{T}$ deviates from $P^{e}_{T}$. 
A natural choice to quantify the contrast between probability densities is the Kullback-Leiber (KL) entropy defined as follows
\begin{equation}
E_{KL}(P,P^{e})=\sum_i P_{T_i}(p) \log(P_{T_i}(p)/P^{e}_{T_i}(p)).
\label{eq9}
\end{equation}
Since the $E_{KL}$ is not a symmetric quantity, we use the following symmetrized form \cite{miami}
\begin{equation}
S_{KL}(p)= \frac{E_{KL}(P,P^{e}) E_{KL}(P^{e},P)}{E_{KL}(P,P^{e})+E_{KL}(P^{e},P)}.
\label{eq10}
\end{equation}
We demonstrated above that order classes are also conjugacy classes. This important property implies that $T$ and $T^{-1}$ belong to the same order class. Thus, $S_{KL}(p)$ for transcriptions inside a class is a suitable invariant measure under the interchange of source and target time series. This property of invariance also allows us to calculate the Kullback-Leiber entropy $S^{\mathcal{C}}_{KL}(p)$ using the probability density of order classes $P_{\mathcal{C}}$ (see Fig. \ref{fig:4}). 
In this case, equations analog to Eqs. (\ref{eq7} - \ref{eq10}) can easily be derived.
\section{Applications}
We apply the method to a bi-directionally coupled Roessler-Roessler system \cite{ls} defined by the following set of equations
\begin{eqnarray}
\dot{x}_{1,2}&=&-w_{1,2}y_{1,2}-z_{1,2}+k(x_{2,1}-x_{1,2}), \nonumber \\
\dot{y}_{1,2}&=&w_{1,2}x_{1,2}+0.165y_{1,2}, \\
\dot{z}_{1,2}&=&0.2+z_{1,2}(x_{1,2}-10).\nonumber
\end{eqnarray}
where $w_{1}=0.99$ and $w_{2}=0.95$ are the mismatch parameters.
All time series were generated using a fourth-order Runge-Kutta method with an increment $\delta t = 0.001$ and the following initial conditions: $x_1 (0)=-0.4$,  $y_1 (0)=0.6$, $z_1 (0)=5.8$, $x_2 (0)=0.8$, $y_2 (0)=-2$, and $z_1 (0)=-4$. Results were saved at intervals $\Delta t =0.01$. 
This chaotic system exhibits a rich synchronization behavior which ranges from phase ($k \approx 0.036$) to lag ($k \approx 0.14$) and finally complete synchronization as the coupling parameter $k$ is increased \cite{ls}.
The results presented here were obtained using the $x$-components of the Roessler subsystems.
Before transforming the time series into symbolic representations, they were
sampled using a sampling time $\tau = 150 \Delta t$ and time series of length $L=2^{19}$ ($\sim$ 775 orbits) were considered. This sampling time fulfills the condition of minimum mutual information of the delay coordinates
$(x_{1}(t),x_{1}(t+\tau))$ for the uncoupled system ($k = 0$) \cite{sw}. Using
this settings, we expect a higher response of our measures to the influence of
the coupling.

\begin{figure}
\centering
\includegraphics[width=14.cm,angle=0]{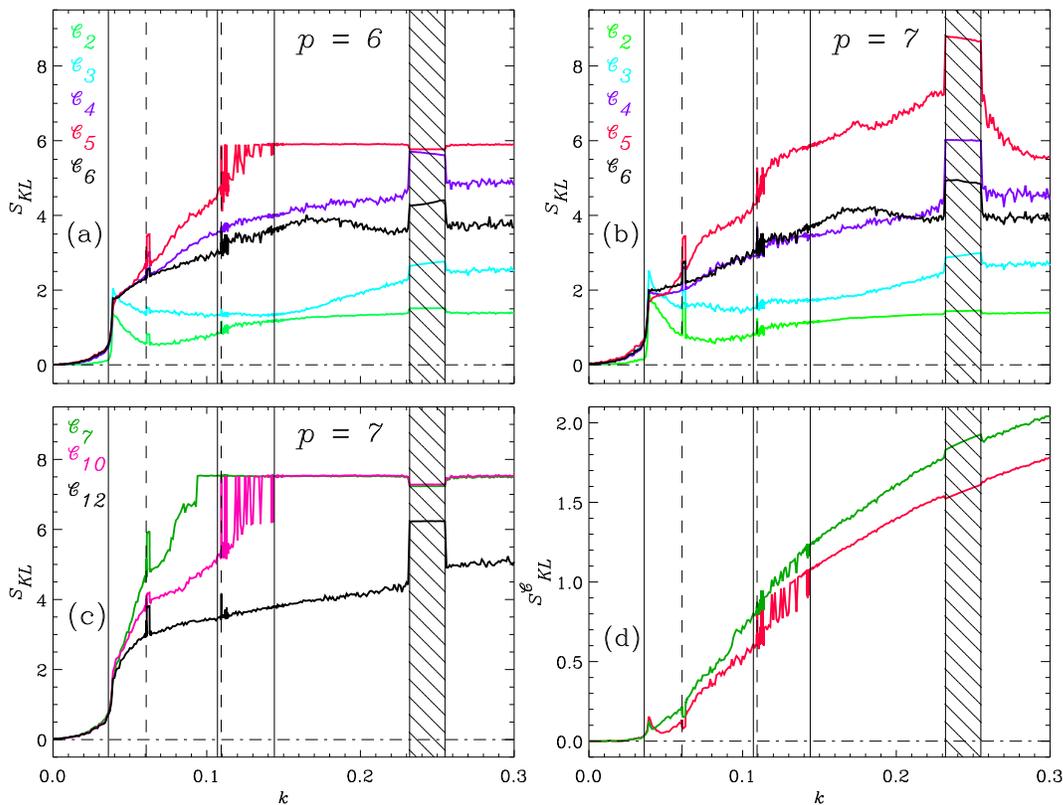}
\caption{(a) Kullback-Leiber entropy $S_{KL}$ obtained using the probability density of transcriptions for all available order classes for $p = 6$. (b) $S_{KL}$ for transcriptions in order classes $\mathcal{C}_2$, $\mathcal{C}_3$, $\mathcal{C}_4$, $\mathcal{C}_{5}$, and $\mathcal{C}_{6}$ for $p = 7$. (c) $S_{KL}$ for transcriptions in order classes $\mathcal{C}_7$, $\mathcal{C}_{10}$, and $\mathcal{C}_{12}$ for $p = 7$. (d) $S^{\mathcal{C}}_{KL}$ obtained using the probability density of order classes for $p = 6$ (red curve) and $p = 7$ (green curve).
Vertical full lines from left to right indicate transitions to phase-,
intermittent lag-, and lag-synchronization, respectively. Vertical dashed lines and hatched areas indicate periodic windows. The values of the coupling constant for transitions and the first periodic window were taken from \cite{ls}.}
\label{fig:3}       
\end{figure}

Figures \ref{fig:3}(a)-(c) show $S_{KL}$ for transcriptions in all feasible order classes for $p=6$ and $p=7$. Figure \ref{fig:3} (d) shows $S^{\mathcal{C}}_{KL}$ obtained using the probability density of order classes $P_{\mathcal{C}}$ for $p=6$ and $p=7$. 
For small values of the coupling constant $k$, the time series behave independently since the Roessler subsystems are uncoupled. For $k \in [0,0.036]$, $S_{KL}$ indicates that the actual dynamics hardly deviates from that of the independent processes.
$S_{KL}$ sharply increases at $k \sim 0.036$ indicating the onset of phase synchronization. At $k \approx 0.061$ all curves display a peak which corresponds to the presence of a period 3 window \cite{ls}. Some curves also indicate the presence of a period 5 window at $k \approx 0.11$. To our best knowledge, the presence of this periodic window has not been reported before probably due to the extremely narrow range of $k$ values ($k \in [0.1094,0.1096]$) where it takes place.
Curves also display a step within the coupling range $k \in [0.232,0.256]$ which indicates the presence of period 5 windows. 
Figures \ref{fig:3}(a) and \ref{fig:3}(c) show that $S_{KL}$ saturates for some order classes. Saturation occurs when an order class 
vanishes, i.e. no transcription belonging to this order class is generated by the coupled dynamics. When this occurs,
the KL entropy is not defined thus Eq. \ref{eq10} can not be used. However, for independent processes the probability density of transcriptions in this particular order class $P_{T}^{e}$ is non-vanishing. In these cases, we found that a reasonable choice is to set $S_{KL}$ to the Shannon information entropy $S_{KL}=-\sum_i P_{T_i}^{e} \log{P_{T_i}^{e}}$ for transcriptions in the vanishing class. 
%

Figure \ref{fig:3} also unveils another interesting feature of this coupled chaotic system. For $k \approx [0.11,0.14]$, $S_{KL}$ displays fluctuations which are particularly strong in Fig. \ref{fig:3}(d) and for some order classes, and sharply decrease for $k > 0.14$. This result provides evidence of the existence of a typical behavior known as intermittent lag synchronization \cite{ls,ils}, 
characterized by synchronization periods interrupted by bursts of non-synchronized behavior. 
These intermittent bursts of activity are responsable for the large fluctuations displayed by $S_{KL}$ in this range of coupling values. The absence of these fluctuations for higher coupling values indicate that bursts of non-synchronized behavior are no longer present thus lag synchronization completely develops. In particular, for $p=6$ ($p=7$) $S_{KL}$ for $\mathcal{C}_5$ ($\mathcal{C}_{10}$), which are the most sensitive measures to this intermittent behavior, saturate at $k \sim 0.14$. This value of the coupling constant is in agreement with the one reported in \cite{ls} ($k = 0.14$) for the onset of lag-synchronization. Figure \ref{fig:3} (d) shows that $S^{\mathcal{C}}_{KL}$ for the probability density of order classes also reveals features above discussed and describes the overall behavior of the coupled system.

\begin{figure}
\centering
\includegraphics[width=15.cm,height=9.4cm,angle=0]{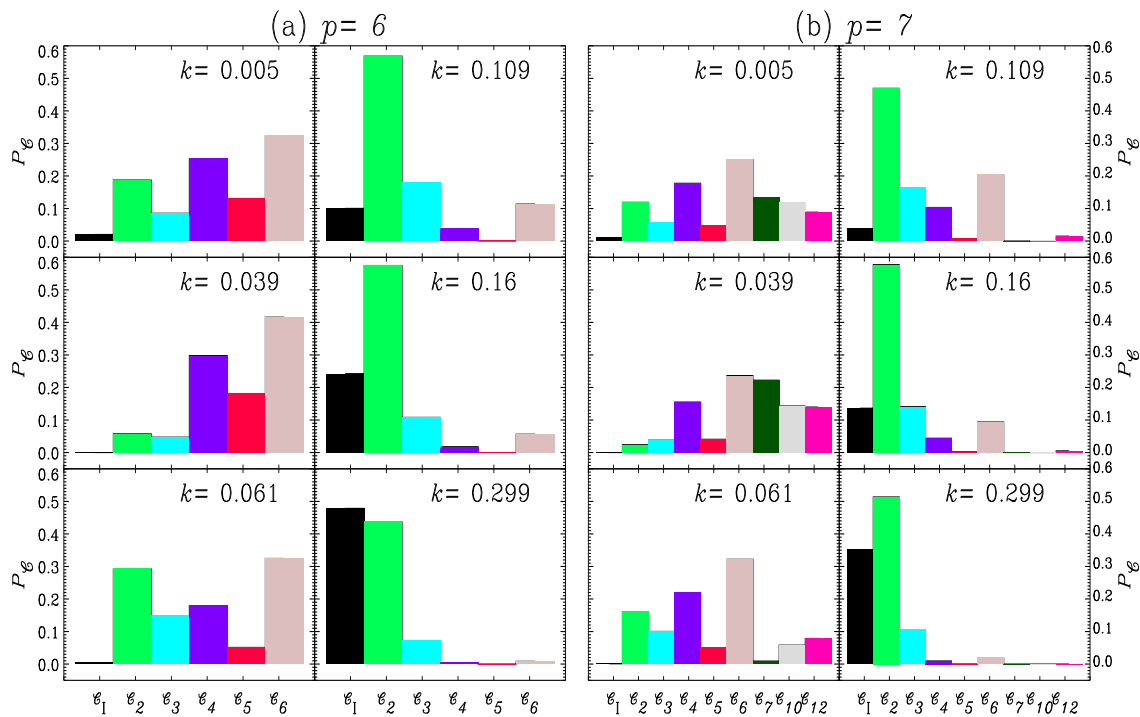}

\caption{(a) Probability density $P_\mathcal{C}$ of the existing order classes for different values of the coupling constant $k$ for $p=6$. Note that class $\mathcal{C}_I$ comprises only one transcription ($I=(0,1,2,3,4,5)$). (b) Idem (a) for $p=7$}
\label{fig:4}       
\end{figure}

Figure \ref{fig:4} (a) and (b) show plots of the
probability density $P_{\mathcal{C}_i}$ of the order classes for $p =6$ and
$p =7$, respectively. Note that Fig \ref{fig:3} (d) shows the contrast between the probability densities shown in Fig. \ref{fig:4} and the ones for independent processes. Figure \ref{fig:3} (d) indicates that for $k = 0.005$ the contrast is vanishing ($S_{KL} \sim 0$) thus
the Roessler subsystems behave independently.
Then, the probability density $P_{\mathcal{C}}$ for $k = 0.005$ is similar
to that generated by two independent processes. 
Note that even for two random independent processes, the probability density of order classes is not
uniform since the cardinality of order classes is different. 
In the vicinity of the transition to phase synchronization, 
$P_{\mathcal{C}}$ deviates from that of the independent processes (see Fig. \ref{fig:3} (d)) and
higher-order classes dominate the coupled dynamics (see Figs. \ref{fig:4} (a)
and (b) for $k = 0.039$). This trend is reversed when increasing $k$ and already
at $k = 0.062$ ($k = 0.074$) for $p=6$ ($p=7$) $\mathcal{C}_2$ is the most
relevant class. 
Figure \ref{fig:4} (a) shows that $\mathcal{C}_2$ dominates up to large values of $k$ where finally $\mathcal{C}_I$ prevails. Figure \ref{fig:4} (b) reveals the same trend as in (a) except that $\mathcal{C}_2$ still dominates at $k=0.299$.  

As discussed above, the order of a transcription roughly estimates its "complexity". Thus, the probability density of order classes indicates how "complex" the relationship between the time series is. Notice that the probability densities of higher-order classes decrease when increasing $k$ and some of them vanish like $\mathcal{C}_5$ for $p=6$, and $\mathcal{C}_7$ and $\mathcal{C}_{10}$ for $p=7$. 
In fact, simpler synchronization states like intermittent lag and lag synchronization ($k > 0.11$) are described by lower order classes ($\mathcal{C}_2$ and $\mathcal{C}_I$). Clearly, the simplest synchronization state, namely complete synchronization, will only be described by $\mathcal{C}_I$. However, for more complex synchronization states like phase synchronization ($k > 0.036$), higher-order transcriptions play an important role. 
\section{Conclusions}
We presented a method to characterize similarities between time series based on symbolic representations which is particularly useful to study synchronization. The properties of invariance that order classes satisfy allow us to derive information measures for the different order classes. Our results show that different order classes provide complementary information of the coupled dynamics.
The understanding of the action of transcriptions belonging to specific order classes led us to interpret the probability density of order classes as an expression of the "complexity" of the existing relationship between the coupled systems. The probability density of order classes shows that more complex synchronization states are mainly described by higher-order classes while lower-order classes dominate for simpler synchronization states.
Our approach to characterize synchronization in time series provides a new frame where elements of group theory and information theory can be directly combined and applied in a simple way. 
We expect our methodology to be useful for the analysis of the dynamics of a wide range of coupled systems, particularly for 
physiological signals like EEG, where the occurrence of synchronization phenomena plays a relevant role.

\ack
We would like to thank Thomas Aschenbrenner for valuable discussions and comments.

\end{document}